 \documentclass[final,5p,times,twocolumn]{elsarticle}

\usepackage{amssymb}
\usepackage{graphics}
\usepackage{graphicx}
\usepackage{epsfig}
\usepackage{threeparttable}
\usepackage{amssymb}
\usepackage{amsthm}
\usepackage{amsmath}
\usepackage[hyphens]{url}
\usepackage{adjustbox}
\usepackage{wasysym}
\usepackage{multirow}
\usepackage{hyperref}% add hypertext capabilities

\usepackage{natbib}

\usepackage{tabularx}
\usepackage{lscape}
\usepackage{pdflscape}
\usepackage{wrapfig}
\usepackage{rotating}
\usepackage{lineno}
\usepackage{fdsymbol}
\usepackage{siunitx}

\usepackage{xcolor}

\biboptions{square,numbers,sort&compress}
\begin{document}

\begin{frontmatter}



\title{Two-step laser resonant ionization spectroscopy of chromium}


\author[inst1,inst2]{Romina Schulz}
\author[inst1,inst3,inst4]{Ruohong Li\corref{cor1}} 
\ead{ruohong@triumf.ca}
\author[inst1,inst2]{Julius Wessolek}
\author[inst1,inst5]{Maryam Mostamand}
\author[inst1,inst6]{Peter Kunz}
\author[inst1,inst5,inst7]{Jens Lassen}

\affiliation[inst1]{organization={TRIUMF-Canada's particle accelerator centre},%Department and Organization
            addressline={4004 Wesbrook Mall}, 
            city={Vancouver}, state={BC},
            postcode={V6T 2A3}, 
            country={Canada}}
            
\affiliation[inst2]{organization={TU Darmstadt},%Department and Organization
            addressline={Karolinenplatz 5}, 
            postcode={D-64285},  city={Darmstadt},
            country={Germany}}

% \author[inst1,inst2]{Author Three}

\affiliation[inst3]{organization={Department of Astronomy and Physics, Saint Mary's University},%Department and Organization
            addressline={923 Robie St.}, 
            city={Halifax}, state={NS},
            postcode={B3H 3C3}, 
            country={Canada}}
            
\affiliation[inst4]{organization={Department of Physics, University of Windsor},%Department and Organization
            addressline={401 Sunset Ave.}, 
            city={Windsor}, state={ON}, 
            postcode={N9B 3P4},  
            country={Canada}}
            
\affiliation[inst5]{organization={Department of Physics and Astronomy, University of Manitoba},%Department and Organization
            addressline={30A Sifton Rd.}, 
            city={Winnipeg}, state={MN},
            postcode={R3T 2N2}, 
            country={Canada}}
\affiliation[inst6]{organization={Department of Chemistry, Simon Fraser University},%Department and Organization
            addressline={8888 University Dr.}, 
            city={Burnaby}, state={BC},
            postcode={V5A 1S6}, 
            country={Canada}}
            
\affiliation[inst7]{organization={Department of Physics, Simon Fraser University},%Department and Organization
            addressline={8888 University Dr.}, 
            city={Burnaby}, state={BC},
            postcode={V5A 1S6}, 
            country={Canada}}
            
\begin{abstract}

At TRIUMF's off-line laser ion source test stand, stepwise resonant laser ionization spectroscopy of chromium (Cr) was carried out, to find an efficient ionization scheme suitable for titanium sapphire (Ti:Sa) laser systems. With three different first-excitation transitions, 357.971~nm, 359.451~nm, and 360.636~nm, automated continuous laser-frequency scans using a frequency-doubled, grating-tuned Ti:Sa laser were performed. Rydberg series as well as autoionizing(AI) states were observed. From these results, the ionization potential (IP) of Cr was determined as 54575.49(2)$_\text{stat}$(2)$_\text{sys}$~cm$^{-1}$, which is one order of magnitude more precise than the previously reported 54575.6(3)~cm$^{-1}$ in NIST database. The ionization scheme using the observed AI resonance, with 357.971~nm as the first step and 373.935~nm as the second step was subsequently deployed to the online delivery of radioactive Cr isotope beams for precision mass measurements. The online yields of $^{50-59}$Cr have been measured at TRIUMF-ISAC.
\end{abstract}


\end{frontmatter}



%\tableofcontents

\section{Introduction}\label{Introduction}

The TRIUMF Isotope Separator and ACcelerator facility ISAC is one of a few radioactive beam user facilities for experimenters in nuclear-, astro-, solid-state- physics as well as nuclear medicine to conduct online radioactive ion beam (RIB) experiments. TRIUMF's ISAC facility uses the isotope separator on-line (ISOL) method to produce isotopes in a thick target, which is irradiated by a continuous 480~MeV proton beam with an intensity of up to 100~µA. This proton beam induces fission, fragmentation, and spallation in the irradiated target material. The target is coupled to an ion source, which allows the ionization, extraction, and acceleration of the reaction products into a fast ion beam for mass separation and prompt delivery to experiments. At ISAC the high-throughput mass-separator magnet can achieve a mass resolution of up to  m/$\Delta$m$\sim$2000. This is sufficient for isotopic separation, but not enough to separate isobars in any but the lightest elements. The resonance ionization laser ion source (RILIS) was implemented at ISAC in 2003~\cite{Jens2005, Jens17} to provide high ionization efficiency and element-selective ionization. The element selectivity of RILIS is based on the unique electronic structures of each element. RILIS enhances the ionization efficiency of the desired element often by orders of magnitude when compared to surface ionization, while leaving the intensity of isobaric contamination virtually unchanged. For operational reasons a Ti:Sa-laser-based RILIS was implemented at TRIUMF, which required that ionization schemes be developed where existing dye-laser-based schemes~\cite{Koe2003} are outside the frequency range of Ti:Sa lasers. The element Cr presents such a case. 

Saloman proposed several possible excitation schemes of Cr in the 1990s~\cite{Saloman1991}. In recent years, several groups experimentally developed RILIS schemes~\cite{Paquet1998, Savina2009, Leven2014, Tom2017}. Among them, Yu et al.~\cite{Leven2014} and Goodacre et al.~\cite{Tom2017} found several efficient schemes with AI states, however, some of the required wavelengths are inaccessible to Ti:Sa lasers. The schemes of Paquet et al.~\cite{Paquet1998} and Levine et al.~\cite{Savina2009} are suitable for Ti:Sa lasers but require three lasers and a high-power, short-pulsed YAG laser for non-resonant ionization. Since Cr has an ionization potential (IP) of about 6.8~eV, a two-step blue-blue scheme will provide enough energy to access autoionizing (AI) resonances energetically above the IP, which could be ideally suited for Ti:Sa-laser-based RILIS operation. To develop such schemes, two-step resonant laser ionization spectroscopy was conducted at TRIUMF’s offline laser ion source test stand (LIS-stand)~\cite{teststandPaper}. A highly efficient two-step blue-blue excitation scheme via an AI state was identified, along with the energies of even-parity high-$n$ Rydberg states not currently listed in the NIST atomic spectra database~\cite{NIST}.

Chromium is a member of the fourth-period transition metals. The Cr spectra are complex as there is competition between the energetically close 3$d$ and 4$s$ shell. However, as a group-six element, its electronic ground state has half-filled 3$d$ and 4$s$ shells, which introduces some simplicity in the optical spectra. In the 1970s Huber et al.~\cite{CurrentIP} observed the odd-parity Rydberg series of chromium, $3d^5(^6S)np$~$^7P_{2,3.4}$, for $n$ = 11-35 using absorption spectroscopy, from which the IP was derived as 54575.6(3)~cm$^{-1}$. This value is still quoted as the most precise value in NIST database. Recently Saloman compiled the reported Cr atomic data~\cite{Saloman2012}. Due to the special characteristics of the Cr electron configuration, most modern spectroscopy studies focus on its inner-shell excitation spectra using synchrotron radiation. The even-parity high-$n$ Rydberg states measured in this work fill gaps in existing data and improve the precision of the ionization potential of Cr~\cite{NIST}.

\begin{figure}[!h]
    \centering
    \includegraphics[width=1\linewidth]{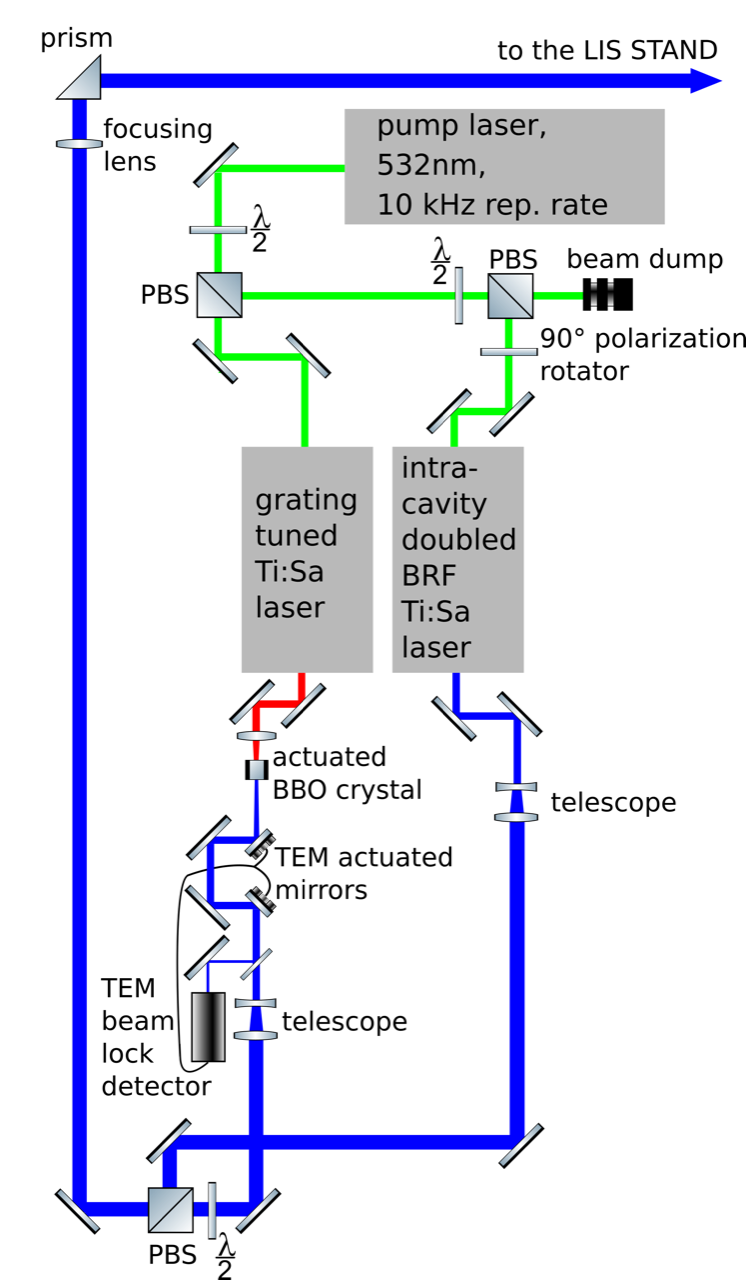}
    \caption{Laser setup for the Cr resonance ionization spectroscopy with two frequency-doubled Ti:Sa lasers. The grating-tuned, externally frequency-doubled Ti:Sa laser was used for the spectroscopy of Rydberg states and systematic search for AI resonances.}
    \label{fig:setup}
\end{figure}

\section{Experimental setup}

The laser system and optical setup for the Cr excitation scheme development is sketched in Fig.~\ref{fig:setup}. Two TRIUMF-built frequency-doubled Ti:Sa lasers were used, pumped by a single multimode frequency-doubled Nd:YAG laser (LEE laser LDP-100MQG), with 160~ns long pulses at 10~kHz repetition rate. The laser for the first excitation step (FES) was birefringent filter/etalon-tuned~\cite{Yi} and intra-cavity frequency doubled; the laser for the second excitation step (SES) was a grating-tuned laser in Littrow configuration~\cite{Andrea_laser_paper, Li_paper_laser} with external frequency doubling. Both laser beams were expanded using a telescope to ensure efficient beam transport and focusing into the ionization region. After expansion, the laser beams were spatially overlapped via a polarizing beam splitter cube (PBS). To match the incident polarization of the PBS, a $\lambda/2$-plate was used to rotate the SES laser polarization. The combined laser beams are steered together by a 50.8 mm diameter broadband dielectric mirror and an uncoated right-angle prism and focused via an uncoated long-focus lens into the ionizer tube located inside the test stand vacuum system, which is $\sim$5~m away from the laser table. To achieve and maintain the temporal overlap of the laser pulses, the lasers were Q-switched by intra-cavity Pockels cells. The pulse timing was manually controlled with a delay gate generator during the scan of the laser wavelength. 

The grating-tuned Ti:Sa laser allowed continuous wavelength scans over a range of 100~nm. To achieve automated second-harmonic scans, a nonlinear LBO crystal was mounted on a computer-controlled rotary stage. The crystal was automatically rotated to maintain phase matching for optimal second-harmonic generation. To maintain spatial overlap of the laser beams in the ionization region, the spatial walk-off of the frequency-doubled laser beam during crystal rotation was compensated by utilizing a commercial beam stabilization system (TEM-Messtechnik BeamLock), as shown in Fig.~\ref{fig:setup}. The TEM Beamlock system monitored both the near-field (translation) and far-field (angle) of a sampled laser beam and controlled two actuated mirrors for laser-pointing correction.  

\begin{figure}
    \centering
    \includegraphics[width=1\linewidth]{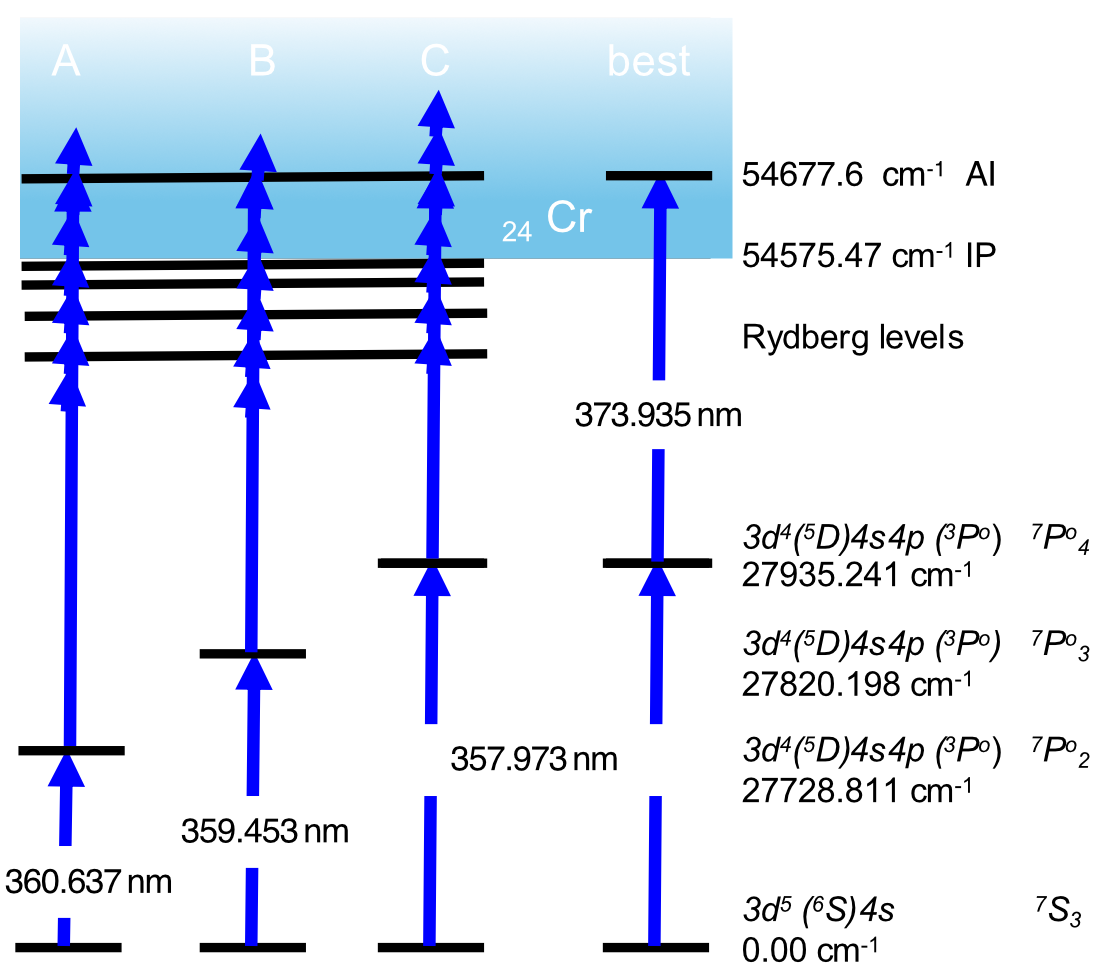}
    \caption{Cr excitation schemes developed at the offline laser ion source test stand. The left side of the plot represents three developed schemes using different first excitation steps. The blue scan covered the region of high-$n$ Rydberg states and ended around $\sim$400~cm$^{-1}$ above the IP. All three scans show a broad AI resonance above the IP. The right side of the plot shows the most efficient scheme identified offline, which was later applied online for radioactive ion beam delivery. All wavelengths here are the values in vacuum.}
    \label{fig:ScanScheme}
\end{figure}

Chromium atomic absorption spectroscopy (AAS) water-based standard solution (Alfa Aesar, 1~µg/µl Cr(NO$_{3}$)$_{3}$ in 5$\%$ HNO$_{3}$)) was loaded quantitatively onto a Ti foil and dried at 110~$^{\circ}$C in an oven. Afterward, the foil was folded and placed inside the Ta ionizer. Inside the test stand vacuum chamber and under the pressure of 10$^{-6}$ Torr, the ionizer can be resistively heated to up to 2000~$^{\circ}$C. To break possible Cr molecular bonds and generate atomic Cr vapor, an ionizer temperature of about 1500~$^{\circ}$C was required. The vaporized atoms were excited/ionized by the lasers and the generated ions were extracted as a beam with 10~keV beam energy for subsequent mass analysis and beam intensity measurement. A quadrupole mass spectrometer (Extrel QMS Max-300) was used to select and detect Cr ions at the mass of interest. To ensure unit mass resolution, the ion beam was decelerated to below $\sim$50~eV beam energy before entering the QMS. Downstream of the QMS a channel electron multiplier (CEM) in counting mode was used to detect the ion signal.

\section{Ionization scheme development using stable Cr isotopes} 
\label{section:Schemedevelopment}

Cr has four naturally occurring isotopes: $^{50}$Cr (4.35$\%$), $^{52}$Cr (83.79$\%$), $^{53}$Cr (9.50$\%$) and $^{54}$Cr(2.36$\%$). For maximum detection sensitivity, $^{52}$Cr was chosen due to its highest natural abundance. Three strong transitions from the Cr ground state to intermediate states accessible via frequency-doubled Ti:Sa laser light were selected as the FESs, as shown in Fig.~\ref{fig:ScanScheme}. Extended wavelength scans from these intermediate states were carried out using the frequency-doubled grating-tuned laser. The scans, spanning 345 to 450~nm, covered the region of high-$n$ Rydberg states and ended $\sim$400~cm$^{-1}$ above the IP. The three intermediate states have the same electronic configuration and terms, $3d^4(^5D)4s4p(^3P^{\circ})$~$^7P^{\circ}$, except for different total angular momenta $J$: scheme A, B, and C are for $J$ = 2, 3 and 4, respectively (Fig.~\ref{fig:ScanScheme}). Consequently, the spectra excited from these three states are similar, with differences arising only from the different $J$, which aids in assigning the observed Rydberg states. 

\begin{figure*}[!h]
    \centering
    \includegraphics[width=1\linewidth]{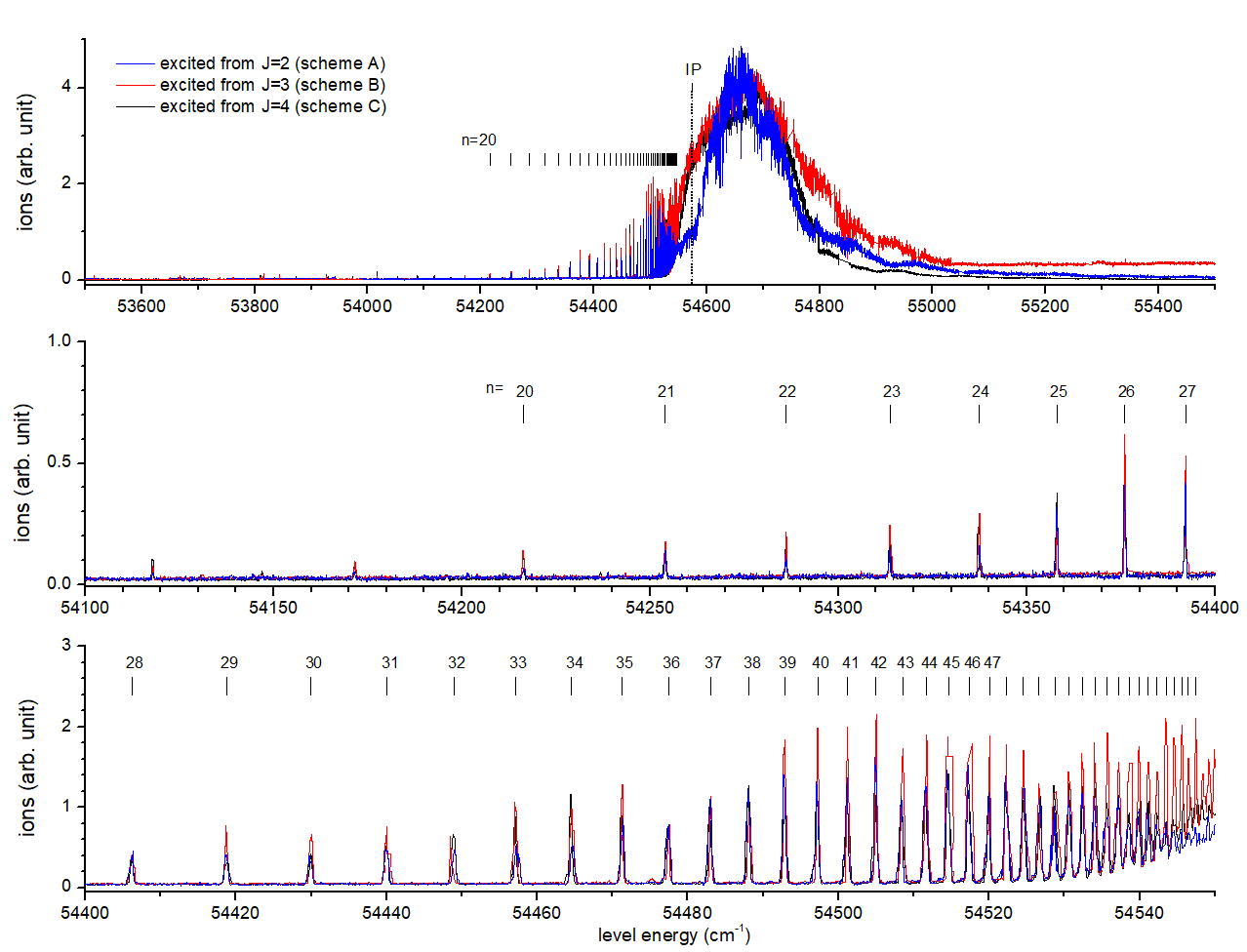}
    \caption{The laser ionization spectra excited from different $J$ states of $3d^4(^5D)4s4p(^3P^{\circ})$ $^7P^\text{o}_{2,3,4}$. a): overview of the complete spectra observed; b) and c): details of Rydberg spectra. The levels of the $3d^5(^6S)ns$~$^7S_3$ Rydberg series are marked with \textquotedblleft$\mid$" and their principal quantum number $n$.}
    \label{fig:ScanPlots}
\end{figure*}

For scheme A and B (Fig.~\ref{fig:ScanScheme}) the heating current of the ionizer tube was kept at 155~A and 150~A, respectively; for scheme C the current was slowly increased from 157~A to 161~A over the span of several hours, to maintain a constant density of Cr atoms in the ionizer throughout the measurement. These heating currents result in a temperature of around 1500~$^{\circ}$C. The recorded spectra are shown in Fig.~\ref{fig:ScanPlots}. Prominent is the broad, high-intensity AI resonance at an energy of 54695~cm$^{-1}$ visible right above the IP. This strong and broad resonance spanning hundreds of cm$^{-1}$, possibly consisting of multiple unresolved AI states, implies a high transition probability from the intermediate state to this AI state, and a strong interaction between the AI state and the underlying continuum (a short lifetime of the AI state). This type of AI state is the ideal target for a highly efficient and robust laser ionization as it provides high ion yields and high tolerance against laser frequency drift. The comparison of ionization efficiency from different intermediate states is presented in Sect.~\ref{Sect:online}.

\section{Rydberg series and ionization potential of Cr}\label{sec:Rydberg series and ionization potential}

An even-parity Rydberg series was observed (Fig.~\ref{fig:ScanPlots}), excited from each intermediate state $3d^4(^5D)4s4p(^3P^{\circ})$~$^7P^\text{o}_{2,3,4}$. The observed series overlap with each other in terms of energy. One possible explanation is that fine structure splittings are not resolved in this high-$n$ range as the laser linewidth of the frequency-doubled grating-tuned Ti:Sa laser is 4-10~GHz. A similar result was previously reported for the series $3d^5(^6S)np$~$^7P^{\circ}_{2,3.4}$ of Cr, where fine structure splittings remained unresolved at the precision of 0.1-0.6~cm$^{-1}$ for $n$ = 11-35~\cite{CurrentIP}. Another possibility is that the observed series has $L$ = 0, and therefore exhibits no fine structures, which was confirmed in subsequent data analysis. The observed spectral peaks were fitted with Gaussian profiles to determine the resonance centroids. The total energies of the Rydberg states were calculated by summing up the intermediate state energy and the SES photon energy. The former was adopted from the NIST data base\cite{NIST}; and the latter was calculated from our experimental laser wavelengths, which were measured by a commercial wavelength meter (HighFinesse WS6-600) with an accuracy of 600~MHz. The wavelength meter was routinely calibrated to a polarization-stabilized HeNe laser (Melles Griot 05 STP 901/903) with a wavelength accuracy of 10$^{-8}$. 

\begin{table}[!h]
	\centering
	\caption{Energies of the Rydberg states $3d^5(^6S)ns$ $^7S_3$, along with their corresponding principal quantum number $n$ and quantum defect $\delta$. The IP value determined in this work (54575.49~cm$^{-1}$) is used to calculate $\delta$ here. The statistic uncertainties $\sigma$ were estimated as the standard error of the three measurements and are presented in parentheses. The systematic uncertainty is 0.02~cm$^{-1}$ from the wavelength meter.}
	\renewcommand{\arraystretch}{1.0}
	\begin{tabular}{cSc}
		\\
		\hline\hline
		$n$ & \multicolumn{1}{c}{energy (cm$^{-1}$)} & $\delta$ \\
		\hline 
		% & & &  \\
	20	&	54216.37(8)	&	2.52	\\
    21	&	54254.00(5)	&	2.52	\\
    22	&	54286.12(4)	&	2.53	\\
    23	&	54313.72(6)	&	2.52	\\
    24	&	54337.42(6)	&	2.53	\\
    25	&	54358.06(5)	&	2.53	\\
    26	&	54376.10(7)	&	2.54	\\
    27	&	54392.18(4)	&	2.53	\\
    28	&	54406.24(4)	&	2.54	\\
    29	&	54418.78(8)	&	2.54	\\
    30	&	54429.96(9)	&	2.54	\\
    31	&	54440.05(5)	&	2.53	\\
    32	&	54449.02(2)	&	2.54	\\
    33	&	54457.19(5)	&	2.54	\\
    34	&	54464.63(3)	&	2.53	\\
    35	&	54471.32(3)	&	2.54	\\
    36	&	54477.44(2)	&	2.54	\\
    37	&	54482.99(5)	&	2.55	\\
    38	&	54488.12(8)	&	2.55	\\
    39	&	54492.87(6)	&	2.55	\\
    40	&	54497.24(4)	&	2.55	\\
    41	&	54501.24(7)	&	2.55	\\
    42	&	54504.99(5)	&	2.54	\\
    43	&	54508.45(7)	&	2.54	\\
    44	&	54511.66(6)	&	2.53	\\
    45	&	54514.61(7)	&	2.54	\\
    46	&	54517.39(7)	&	2.53	\\
    47	&	54520.02(4)	&	2.51	\\
    48	&	54522.34(6)	&	2.55	\\
    49	&	54524.61(5)	&	2.55	\\
    50	&	54526.74(11)	&	2.54	\\
    51	&	54528.77(6)	&	2.53	\\
    52	&	54530.66(6)	&	2.51	\\
    53	&	54532.46(7)	&	2.49	\\
    54	&	54534.08(8)	&	2.51	\\
    55	&	54535.67(8)	&	2.49	\\
    56	&	54537.23(5)	&	2.43	\\
    57	&	54538.63(8)	&	2.42	\\
    58	&	54539.94(8)	&	2.43	\\
    59	&	54541.18(7)	&	2.43	\\
    60	&	54542.39(7)	&	2.40	\\
    61	&	54543.54(9)	&	2.38	\\
    62	&	54544.68(7)	&	2.30	\\
		\hline 
		\hline 
	\end{tabular}
	\label{tab:RydbergStates}
\end{table}

\begin{figure}[!h]
    \centering
    \includegraphics[width=1\linewidth]{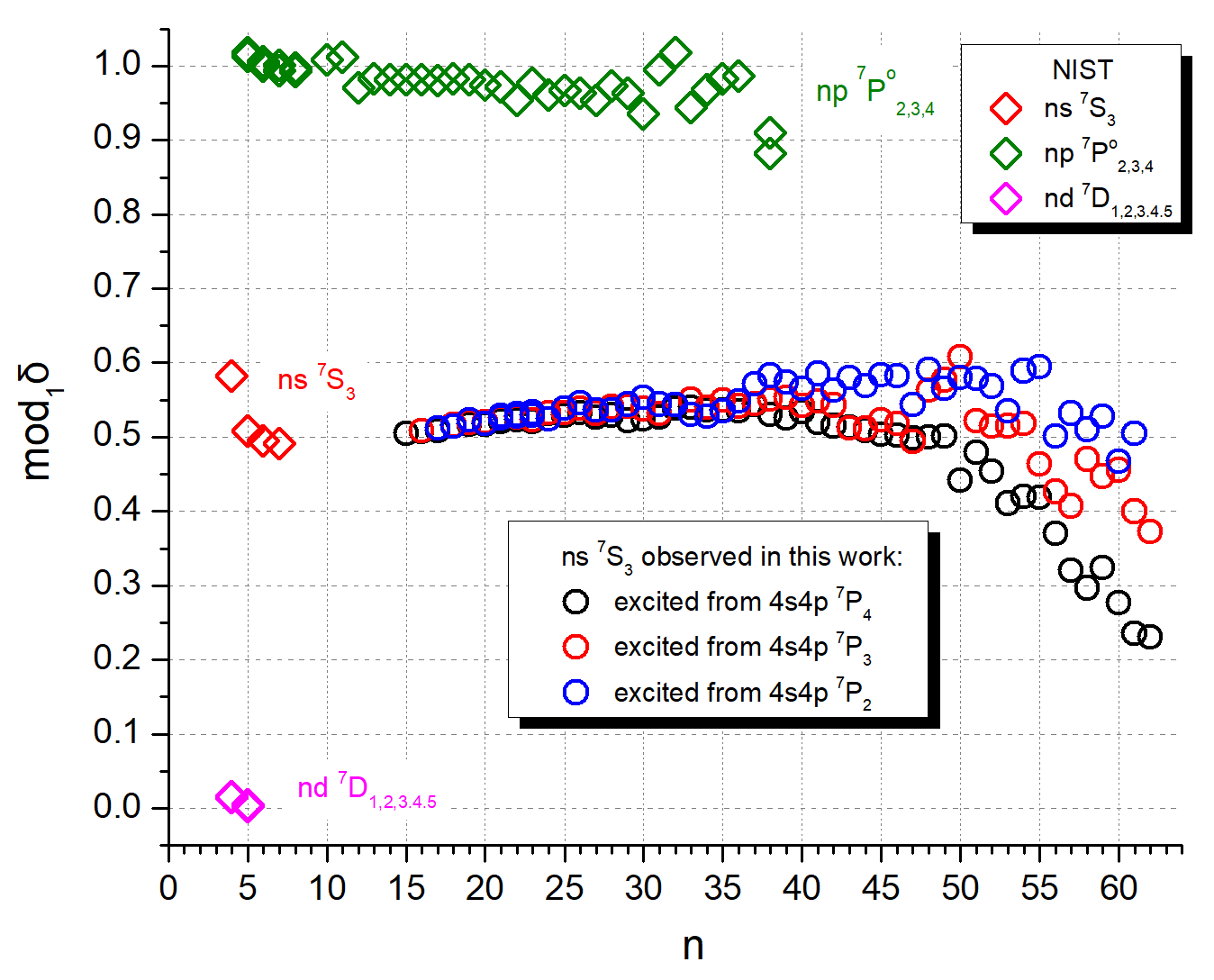}
    \caption{Lu-Fano plot of the observed Rydberg series $3d^5(^6S)ns$ $^7S_3$, with comparison to previously reported data of $3d^5(^6S)ns$ $^7S_3$, $3d^5(^6S)nd$ $^7D_{1,2,3,4,5}$ and $3d^5(^6S)np$ $^7P_{2.3.4}$ from the NIST atomic database~\cite{NIST}. Huber et al.~\cite{CurrentIP} used the series $3d^5(^6S)np$ $^7P^{\circ}_{2.3.4}$ to determine the IP value to be 54575.6(3)~cm$^{-1}$, which is the value currently adopted by NIST. The quantum defects $\delta$ plotted here are calculated using the IP value determined in this work (54575.49(2)$_\text{stat}$(2)$_\text{sys}$~cm$^{-1}$).}
    \label{fig:LuPlot}
\end{figure}

\begin{figure}[!h]
    \centering
    \includegraphics[width=1\linewidth]{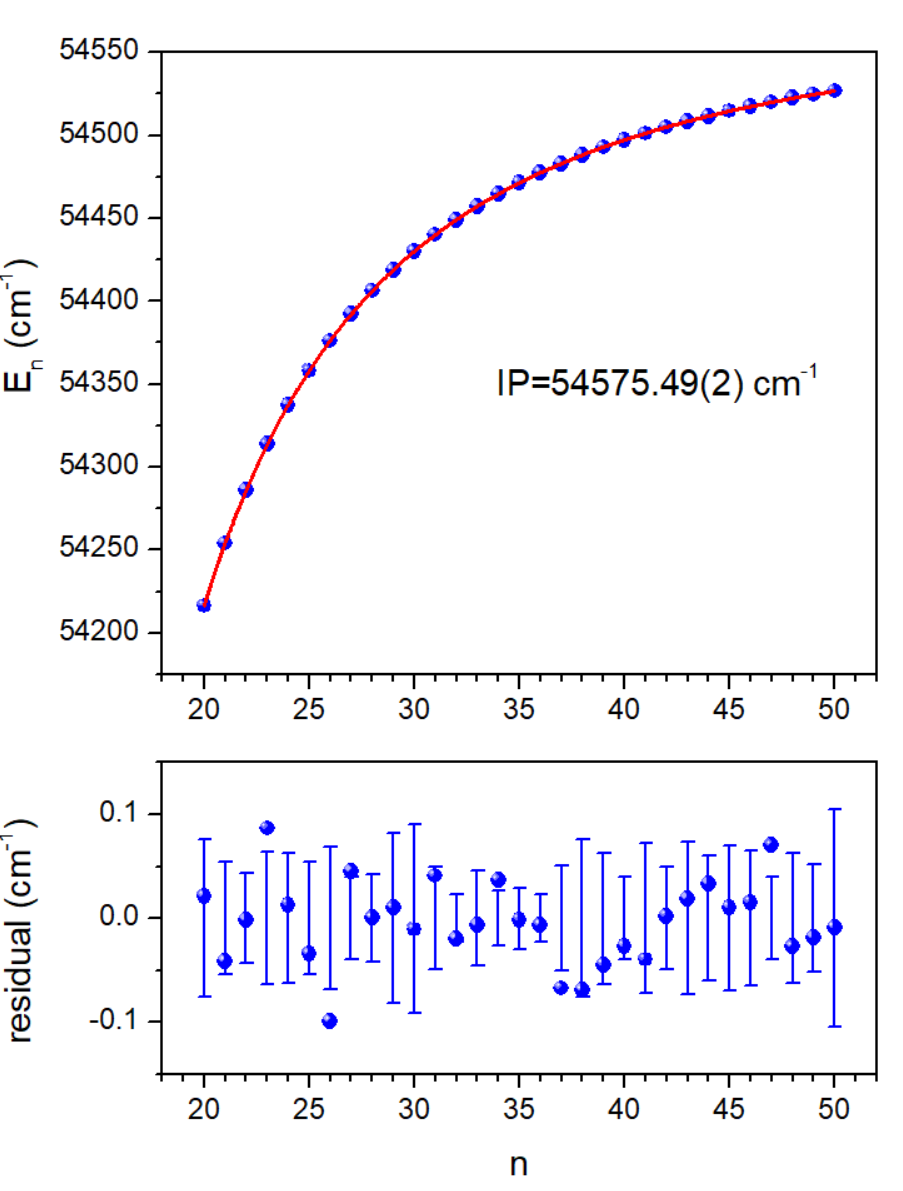}
    \caption{Rydberg-Ritz fit of the Rydberg series $3d^5(^6S)ns$ $^7S_3$ for $n$ = 20-45 to extract the IP value. The two terms in the Ritz expansion were utilized as shown in Eq. \ref{Ritz_2}. The standard errors $\sigma$ of the measurements from three different spectra, which are too small to see in the $E_n$ plot, are plotted with the residual.}
    \label{fig:IPFit}
\end{figure}

To clearly identify the observed Rydberg series, a Lu-Fano plot was constructed (Fig.~\ref{fig:LuPlot}). The quantum defect $\delta$ remains approximately constant with $\delta$mod$_1$ $\approx$ 0.5 from $n$ = 19 up to 45, but begins to deviate as the Rydberg series approaches the AI state, located at $\sim$120~cm$^{-1}$ above the IP with its wing onset around $n$ = 50 beneath the IP. This implies strong configuration interaction between the Rydberg series and the AI state, consistent with the observed broad profile of the AI state embedded in the continuum toward which the Rydberg series converges. Excited from the intermediate state of $3d^4(^5D)4s4p(^3P^{\circ})$ $^7P^\text{o}_{2,3,4}$, the Rydberg series should be $3d^5(^6S)ns$ or $n$d. From available atomic data in NIST database, $3d^5(^6S)ns$ has $\delta$mod$_1$ $\approx$ 0.5, and $3d^5(^6S)nd$ has $\delta$mod$_1$ $\approx$ 0.0, as shown in Fig.~\ref{fig:LuPlot}. Therefore the observed Rydberg series should be $3d^5(^6S)ns$ $^7S_3$. This also explains why only one series was observed by excitation from different $J$ states. Since the same series was measured in the three different spectra, the mean values are calculated and presented in Tab.~\ref{tab:RydbergStates}. The standard error of the mean $\sigma$ of the measurements from three different spectra, ranging from 0.02 to 0.09~cm$^{-1}$ (except the data at $n$ = 50 with $\sigma$ = 0.11~cm$^{-1}$), reasonably falls within the typical uncertainty of 0.15~cm$^{-1}$ of the measured Rydberg-state energies at our test stand. This 0.15~cm$^{-1}$ uncertainty was estimated by comparing the previously measured Rydberg-state energies of Sb, Lu, and Te at TRILIS test stand~\cite{Sb, Lu, Te} with NIST data~\cite{NIST}. The statistic uncertainty $\sigma$ is presented in parenthesis with $E_n$ in the Tab.~\ref{tab:RydbergStates}. The uncertainties of the three intermediate excited states are 0.0019 to 0.0020~cm$^{-1}$~\cite{NIST}, which are small and negligible relative to our uncertainty. The systematic uncertainty from the wavelength meter (HighFinesse WS6–600) is 600~MHz, i.e. 0.02~cm$^{-1}$, according to a 3-$\sigma$ criterion~\cite{WS6}.

A precise value of the IP can be extracted from Rydberg series $3d^5(^6S)ns$ $^7S_3$ by using the Rydberg-Ritz formula:
\begin{equation}
E_{n}=\text{IP}-\frac{R_\text{M}}{n^{*2}}=\text{IP}-\frac{R_\text{M}}{(n-\delta(n))^2},
\label{Ritz_1}
\end{equation}
\begin{equation}
\delta(n)=\delta_{0}+\frac{a}{(n-\delta_{0})^2}+\cdots
\label{Ritz_2}
\end{equation}

Here $E_n$ is the energy of the Rydberg state with the principal quantum number $n$, \text{IP} is the ionization potential, $R_{M}$ is the mass-corrected Rydberg constant (109736.157~cm$^{-1}$ for $^{52}$Cr), and $\delta(n)$ is the quantum defect. Eq. (2) is the Ritz expansion of $\delta(n)$, in which $\delta_0$ is the asymptotic limit of the quantum defect, and $a$ is the second-order expansion coefficient. To avoid the perturbation effect from the AI state, only the states of $n$ = 20-50 were used in the data analysis. The weight applied for fitting was the $1/\sigma^2$. The fitted curve and corresponding residuals are shown in Fig.~\ref{fig:IPFit}. The error bars of $E_n$ ($\pm\sigma$) are plotted with the residuals in Fig.~\ref{fig:IPFit}, as they are too small to be visible in the $E_n$ plot. Using only the first term $\delta_0$ in Ritz expansion results in a systematic deviation of the residuals from zero at $n$ = 20-25 region, reflecting the core effect at low-energy Rydberg states. Adding the second-order expansion term solved this problem, and no obvious further improvement was found by including higher orders. The final fit shown in Fig.~\ref{fig:IPFit} uses two expansion terms. The residuals are evenly distributed around zero and within $\pm$ 0.10~cm$^{-1}$. The extracted IP is 54575.49~cm$^{-1}$ with a fitting error of 0.02~cm$^{-1}$. The error here has been scaled by reduced chi-square $\chi_r^2$ = 0.6315. The resulting $\chi_r^2$, which is smaller than 1, indicates the statistical uncertainty $\sigma$ of $E_n$ (listed in Tab.~\ref{tab:RydbergStates}) are slightly overestimated. The fitted value of quantum defect $\delta$ is 2.552(4). The systematic uncertainty on the IP value is estimated as 0.02~cm$^{-1}$ as the same as the Rydberg-state energies. The determined IP value of 54575.49(2)$_\text{stat}$(2)$_\text{sys}$~cm$^{-1}$ improves the precision, compared to the currently accepted IP value of 54575.6(3)~cm$^{-1}$ measured by Huber et al.~\cite{CurrentIP} more than 40 years ago. The odd-parity Rydberg series $3d^5(^6S)np$~$^7P^{\circ}_{2.3.4}$, which Huber et al. used to extract the IP, was also plotted in the Lu-Fano plot (green diamond in Fig.~\ref{fig:LuPlot}). The constant quantum defects of these data show their consistency with the improved IP value.

\section{Radioactive Cr beams development}
\label{Sect:online}

The Cr RILIS scheme development described in this paper was initiated by a request of TRIUMF's Ion Trap for Atomic and Nuclear Science (TITAN) group for neutron-rich Cr isotopes for high-precision mass measurements. Such mass measurements give insight into the neutron binding energies and nuclear structure for $^{62-65}$Cr~\cite{TITAN}. In October 2019, the developed Cr scheme was applied online at ISAC for mass measurement experiments, and radioactive isotope yields were measured. During the beam delivery, a 15~µA 480~MeV proton beam impinged on a uranium carbide (UCx) target with a Re-lined hot-cavity ionizer tube. The target temperature was estimated to be 1930~$^{\circ}$C as derived from the calculated power deposition from the proton beam and the resistive heating of the target ion source. The hot-cavity ionizer was operated at the typical current 230~A, which provides a temperature around 2000~$^{\circ}$C at the laser ionization region. The ionization energy of Cr is about 6.8~eV, resulting in a lower surface ionization efficiency. The surface ionized Cr background can be neglected in the yield measurement.  

\subsection{Cr resonance ionization scheme selection}

With all three investigated schemes exciting to the same AI state, one still needs to determine the most efficient scheme for online operation. To minimize the influences of atom-vapor-density fluctuations and other systematic long-term drifts, ion counts obtained from the three schemes were measured offline within 30 minutes. These measurements were conducted with a well-stabilized and thermalized system. The ionizer current was kept constantly at 160 A, corresponding to a temperature of 1500~$^{\circ}$C. The relative ion yields are listed in Tab.~\ref{tab:Intensity}.

\begin{table}[h]
	\centering
	\caption{Off-line comparison of the efficiency of the three possible schemes. All presented wavelengths $\lambda$ are vacuum wavelengths, and $P$ are the laser powers used in measurements. The FESs are all saturated at laser power 0f 10-20~mW, therefore well saturated at the used laser power in the comparison measurement. }
	\renewcommand{\arraystretch}{1.2}
		\begin{adjustbox}{width=0.45\textwidth}	
	\begin{tabular}{ccccccccc}
		\\
		\hline \hline
		&&\multicolumn{2}{c}{FES}&&\multicolumn{2}{c}{SES}&& ion yield \\
		\cline{3-4}\cline{6-7}	\cline{9-9}
		scheme&&$\lambda_{vac}$&$P$&&$\lambda_{vac}$&$P$&&\\
		&&(nm)&(mW)&&(nm)&(mW)&&(arb. unit)\\
	  \cline{3-4}\cline{6-7}	\cline{9-9}
		A&&360.636 &137 &&371.375 &106 &&0.75 $\pm$ 0.05\\
		B&&359.451 &176 &&372.182 &90 &&1.4 $\pm$ 0.1\\
		C&&357.971 &166 &&373.935 &110 &&1.9 $\pm$ 0.1\\		
		\hline 
		\hline 	
	\end{tabular}
    \end{adjustbox}
	\label{tab:Intensity}
\end{table}

The laser power was measured between the beam-expansion telescopes and the polarizing beam splitter (Fig.\ref{fig:setup}). To estimate the laser light density inside the 3-mm diameter ionizer tube, which is 5~m away from the laser table, a power loss of approximately 50$\%$ during transport must be taken into account. The powers of the FES lasers were well above the saturation power of the corresponding transitions, which were observed to saturate at 10-20~mW under the present experimental conditions. Therefore, the $<$10\% variation in laser power resulting from wavelength change between the different FESs does not impact the outcome of the comparison. For all three schemes, the SES to the AI state exhibits a linear dependence on laser power up to the maximum power available at the time. Since the laser powers used for all three SESs were nearly identical, the observed ion yields effectively represent the relative ionization efficiencies. Among them, Scheme C, 357.971~nm (FES) + 373.935~nm (SES), shows the highest efficiency.

\begin{figure}[!ht]
    \centering
    \includegraphics[width=1\linewidth]{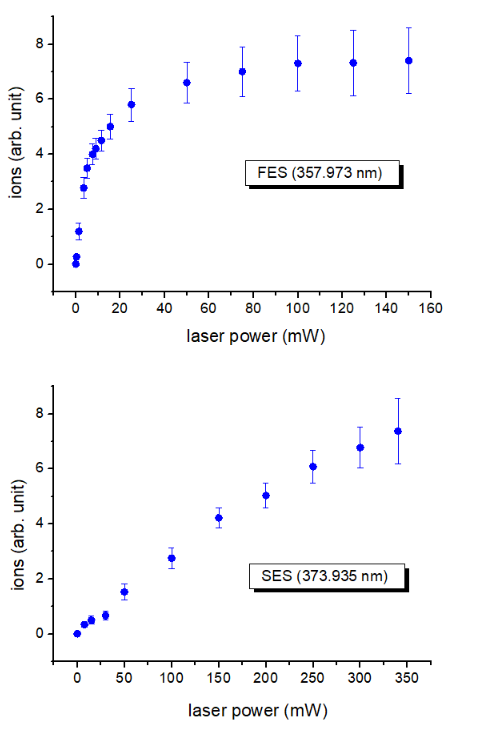}
    \caption{The saturation curves of two excitation steps of the most efficient ionization scheme (357.973~nm + 373.935~nm), measured using $^{52}$Cr during online beam delivery at ISAC in Oct 2019. All wavelengths given are the values in vacuum. The FES saturates at around 7~mW, while the SES has a nearly linear dependence on laser power up to 350~mW. For online beam delivery, birefringent filter/etalon-tuned Ti:Sa lasers with intra-cavity frequency doubling are used for both steps, which by now routinely deliver more than 750~mW.}
    \label{fig:Saturation}
\end{figure}

\subsection{Saturation measurement on radioactive Cr}

During the online run, saturation curves were remeasured with stable $^{52}$Cr, using a Faraday cup with a current integrator downstream from the ISAC high-resolution mass separator. To eliminate isobaric contamination and background current on the Faraday cup, the ion signal difference between laser on and off was recorded. The laser power was measured using the same method as in the offline measurements and with a similar laser transport power loss ($\sim$ 50$\%$) between the measurement spot and the ionization region inside the target chamber. Since the online laser system could provide a higher laser power for the SES, it gave a wider laser power range to investigate the saturation behavior. As seen in the resulting plot (Fig.~\ref{fig:Saturation}), the saturation curve of the SES is still linear at the highest accessible laser power of 340~mW. This implies a higher ionization efficiency can be expected with improved laser power in the SES. The FES transition reached saturation at about 7~mW, as shown in Fig.~\ref{fig:Saturation}.

\subsection{Cr isotope yields}

\begin{figure}[!h]
    \centering
    \includegraphics[width=1\linewidth]{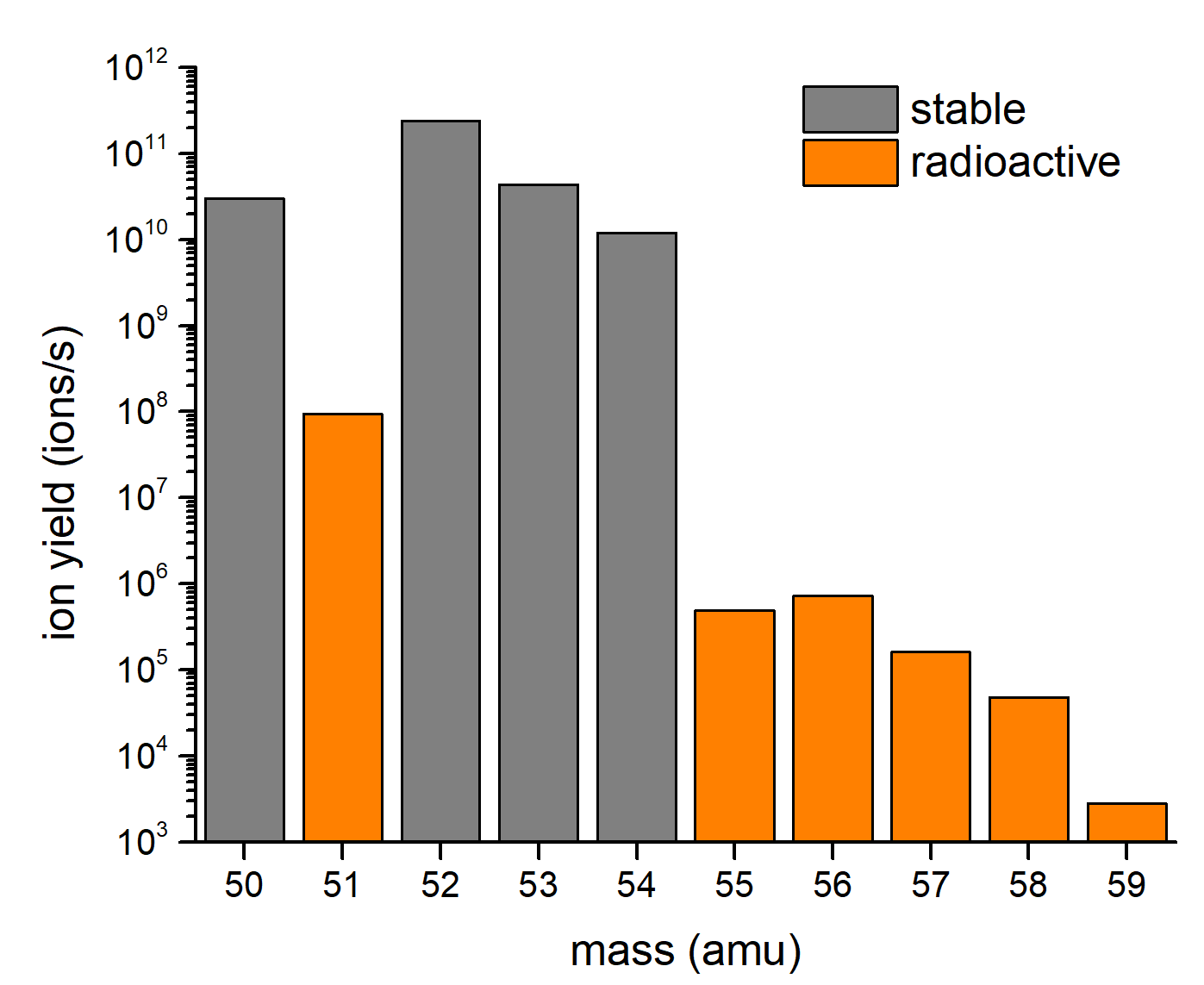}
    \caption{Logarithmic plot of the Cr isotope yields (ions/s) during the online beam delivery in October 2019 at ISAC, TRIUMF~\cite{yield_database_peter}.}
    \label{fig:Yield}
\end{figure}

Yield measurements of Cr isotopes were performed using the scheme 357.971~nm (FES) + 373.935~nm (SES), and the maximum accessible laser powers: 150~mW for the FES and 340~mW for the SES. The maximum accessible power refers to the conservatively available output, ensuring stable laser frequency and power for reliable multi-day operation (as of 2018). For stable and long-lived isotopes $^{50-54}$Cr, yields were measured as the difference in ion signal with the laser on and off using a Faraday cup. For the short-lived radioactive isotopes $^{55-59}$Cr, yields were determined via beta-decay counting of the ion beam implanted in the ISAC yield station~\cite{peter_yield_paper}. The obtained isotope yields are presented in Fig.~\ref{fig:Yield} and Tab.~\ref{table_yield}. The yields for stable Cr isotopes were significantly higher than those for the radioactive isotopes. The observed difference exceeds the expected in-target build-up of long-lived isotopes, as the long-lived radioactive isotope $^{51}$Cr (t$_{1/2}$=27.7 days) also has a yield of 2-3 orders of magnitude lower than the stable isotope yields. Our working hypothesis is that the stable Cr excess stems from Cr contamination either from the UCx target material and material processing, the materials of the target container and ionizer tube (Ta and Re), or from incursions during machining processes.

\begin{table} [h!] \footnotesize
	\caption{Online Cr isotope yields from a UCx target with a Re-lined hot-cavity ion source coupled with laser ionization. The target was at 1930~$^{\circ}$C and the hot-cavity ionizer tube was at 2000~$^{\circ}$C. The ionization scheme, 357.971~nm (FES) + 373.935~nm (SES), was used with laser powers of 150~mW for the FES and 340~mW for the SES.}
	\begin{center}
		\begin{threeparttable}
			
			%\begin{adjustbox}{width=0.5\textwidth}
			\renewcommand{\arraystretch}{1.4}
			\begin{tabular}{cccc}
				\hline\hline
				isotope	&	measurement method & half life	& yield (ions/s)\\			
				\hline											
				$^{50}$Cr	&	Faraday cup	&	stable	&	3.0$\times$10$^{10}$\\
				$^{51}$Cr	&	Faraday cup	&	27.7~d	&	9.4$\times$10$^{7}$	\\
				$^{52}$Cr	&	Faraday cup	&	stable	&	2.4$\times$10$^{11}$\\
				$^{53}$Cr	&	Faraday cup	&	stable	&	4.4$\times$10$^{10}$\\
				$^{54}$Cr	&	Faraday cup	&	stable	&	1.2$\times$10$^{10}$\\
				$^{55}$Cr	&	beta decay	&	3.497~m &	4.9$\times$10$^{5}$	\\
				$^{56}$Cr	&	beta decay	&	5.94~m	&	7.3$\times$10$^{5}$	\\
				$^{57}$Cr	&	beta decay	&	21.1~s	&	1.6$\times$10$^{5}$	\\
				$^{58}$Cr	&	beta decay	&	7.9~s	    &	4.8$\times$10$^{4}$ \\
				$^{59}$Cr	&	beta decay	&	0.74~s	&	2.8$\times$10$^{3}$	\\
				\hline\hline
			\end{tabular}
			%\end{adjustbox}
		\end{threeparttable}
	\end{center}
	\label{table_yield}
\end{table} 

\section{Conclusion}

A two-step resonant ionization scheme of chromium was developed at TRIUMF's offline LIS stand: the first step is to excite the Cr atoms from the ground state $3d^5(^6S)4s$~$^7S_3$ to the intermediate state of $3d^4(^5D)4s4p(^3P^\text{o})$ $^7P^{\circ}_{2,3,4}$; and the second step is to excite the Cr atoms to further higher energy levels by scanning the laser frequency across the high-$n$ Rydberg region, the IP and beyond, to obtain the laser ionization spectra and search for AI states. An even-parity Rydberg series $3d^5(^6S)ns$ $^7S_3$ and a broad AI state at 54695~cm$^{-1}$ were observed. The AI state provides an efficient laser ionization scheme, which was applied to online Cr radioactive ion beam delivery at TRIUMF-ISAC. The ion yields of Cr isotopes were measured and the saturation behaviors of the two laser excitation steps with respect to laser power were investigated. From the observed Rydberg series $3d^5(^6S)ns$ $^7S_3$, the IP was determined as 54575.49(2)$_\text{stat}$(2)$_\text{sys}$~cm$^{-1}$. This represents an order of magnitude improvement over the previously reported first ionization potential~\cite{CurrentIP}. 

\section*{Acknowledgements}

The experimental work is funded by TRIUMF which receives federal funding via a contribution agreement with the National Research Council of Canada and through a Natural Sciences and Engineering Research Council of Canada individual Discovery Grant NSERC RGP-IN-2017-00039 to J.~Lassen.

%\appendix

%\section{Sample Appendix Section}
%Lorem ipsum dolor sit amet, consectetur adipiscing elit, sed do eiusmod tempor section \ref{sec:sample1} incididunt ut labore et dolore magna aliqua. Ut enim ad minim veniam, quis nostrud exercitation ullamco laboris nisi ut aliquip ex ea commodo consequat. Duis aute irure dolor in reprehenderit in voluptate velit esse cillum dolore eu fugiat nulla pariatur. Excepteur sint occaecat cupidatat non proident, sunt in culpa qui officia deserunt mollit anim id est laborum.

% \bibliographystyle{elsarticle-num} 
% \bibliography{cas-refs}


\end{document}